# MULTI-AGENT BASED IOT SMART WASTE MONITORING AND COLLECTION ARCHITECTURE

Eunice David Likotiko, Devotha Nyambo, Joseph Mwangoka

Information and Communication Science and Engineering, The Nelson Mandela African Institution of Science and Technology (NM-AIST), Tanzania.

## *ABSTRACT*

*Solid waste management is one of the existing challenges in urban areas and it is becoming a critical issue due to rapid increase in population. Appropriate solid waste management systems are important for improving the environment and the well-being of residents. In this paper, an Internet of Things(IoT) architecture for real time waste monitoring and collection has been proposed; able to improve and optimize solid waste collection in a city. Netlogo Multi-agent platform has been used to simulate real time monitoring and smart decisions on waste management. Waste filling level in bins and truck collection process are abstracted to a multi-agent model and citizen are involved by paying the price for waste collection services. Furthermore, waste level data are updated and recorded continuously and are provided to decision algorithms to determine the vehicle optimal route for waste collection to the distributed bins in the city. Several simulation cases executed and results validated. The presented solution gives substantial benefits to all waste stakeholders by enabling the waste collection process to be more efficient*

## 1. INTRODUCTION

The amount of waste generated in urban areas is proportional to the rapid growth of cities. The increase of human activities and consumption patterns generate various types of waste that must be appropriately managed to ensure sustainable development and a decent standard of living for all urban residents. Managing Urban solid waste is complex especially to most of the resourced constrained regions. Conventional approach to Municipal Solid Waste Management (MSWM) suffers from lack of sustainable collection throughput [1], lack of data on collection time and location, lack of efficient monitoring systems of waste bin status, lack of efficient systems for waste collection and transportation, and delays in collection. Creativity of having reliable approach on solid waste management is needed, starting from the existing strengths of the city and build up on them with good involvement of all stakeholders in designing their own local models [2]. The use of technology can be adopted to improve the quality of waste data collection, the service availability and reliability in a city thus, building smart cities' world in the aspects of waste management.

According to [3]"A Smart City is a city well performing in a forward-looking way in the following fundamental components (i.e., Smart Economy, Smart Mobility, Smart Environment, Smart People, Smart Living, and Smart Governance), built on the 'smart' combination of endowments and activities of self-decisive, independent and aware citizens". Therefore, smart city process for waste collection is a fundamental point in achieving green city environment and well-being of the citizen and its quality should be considered seriously. smart home applications are of trending.

In this scenario, the application of Information and Communications Technologies (ICT) solutions through Internet of Things(IoT) is of interest.[3]and[4]defined IoT as: The Internet of Things allows people and things to be connected anytime, anywhere (AAA), with anything and anyone, ideally using any path/network and any service. The IoT concept enables easy access and interaction with a wide variety of devices such as, home appliances, surveillance cameras, monitoring sensors, actuators, displays, vehicles, and so on[5]. The idea of internet of things (IoT) was developed in parallel to Wireless sensor networks (WSNs). WSNs as an important aspect in IoT, changes and advances environmental monitoring scenarios and increasingly being adopted in different application[6]for instance, waste collection and monitoring activities. The real timewaste monitoring is now possible from the IoT enabled bins which can sense the status of waste level in a bin and update the status to the central server through the connections. In addition, citizens are now given a room to fully participate in the process of waste management, as a way to build a smart city in the aspect of smart waste management.

Multi-agent based models presents tools and methods to visualize and abstract a given system. Multi-agent system (MAS) "is a set of software agents that interact to solve problems that are beyond the individual capacities or knowledge of each individual agent"[7].Many researches have been done using the agents in providing simulated solutions for different smart city scenarios, for instance intelligent waste collection[8], and generic model for smart city[9]. In this paper, a Multiagent based simulation model is used to simulate the functioning of the IoT and smart city in a waste management scenario.

The aim of this work is to abstract real time solid waste monitoring and collections using multi-agent based model, to design architectural models for real time solid waste's bin monitoring and collection based on wireless sensor network (WSN) technology. WSN can be useful in smart waste management, to overcome the challenges on MSWM, providing optimal path for waste collection resulting into reduced operational costs. Furthermore, this paper highlights the importance of information flow between the authorities and waste collection points with a target of increasing collection throughput, optimize routes and continuously maintaining statistical records of the amount of waste collected, which are necessary for the sustainability of the operations.

The remaining part of this paper has been divided into six sections: Related work, Methodology used with the study requirement analysis and Designs, Simulation set up, Results Discussion, Conclusion Future work and Acknowledgment.

## 2. Related Work

Several efforts have been invested in tackling the problem of Municipal Solid Waste Management (MSWM). Most of the initiatives use wireless communication technologies as one of the solution in reducing amount of time and cost for waste collection and transportation and making smart waste management.

A real time intelligent bin status monitoring and rule based decision algorithms[1] is among the invested efforts toward smart solid waste management. The monitoring application is based on decision algorithms for sensing solid waste data via a wireless sensor network. The system is built on a three-level architecture; smart bin, gateway and control station.

The elementary concept is that, smart bins collect their status when any changes occur and transmit the status data to a server via an intermediate coordinator. A set of applications in server presents the updated bin status on real time. The strength of the system is the design and development of an automated bin status monitoring system that exercises a set of novel rules based decision algorithms. However, the system encounters technical challenges technical such as long range communication technologies on gateways, erroneous output of data from sensors readings, and lack of GPS for location detection, citizen involvement in the system for better interactions which is proposed by this paper.

Automation and integrated system by[10]performed 55 tests run on the developed prototype. The authors provide the unique effort on waste monitoring by incorporating accelerometer, magnetic proximity, ultrasonic, weight sensing and long lived rechargeable battery respectively. The integrated sensing system is designed using rule-based decision procedure to offer a proficient and automatic bin status monitoring system. However, the designed system does not support the use of the obtained waste data for route optimization and how the citizen as the major waste stakeholders will play their role in making sure the smart waste management is achieved.

Similarly,[11]used an open-source project Smart-M3 Platform which has smart-space for virtualization of the real environment used in simulating the real time waste collection. All components of the system; Proximity and Weight sensors, Raspberry PI, Xbee module were implemented using Python language and the ontology was developed using Protege. Authors argued that several simulation cases were run and a map of full and semi full bin was developed and the vehicle was sent to them. In additional a user android application integrated with the Google Map API was still to be a part of the implementation. Besides of the simulated tests the proposed approach didn't consider the development of the central system for users and their bins registration and authenticity also the graphical user interface for central system. Moreover, the system lacks the route decision making algorithm of vehicle for waste collection In [12]used RFID, cell load sensors technologies and Personal digital assistant (PDA) in the achievement of smart waste management. Authors detailed that RFID tag are embedded on the bin, the PDA (smart phone) consists of a reader which converts the radio waves reflected from the waste bin into digital information (i.e., bin ID) that is then recorded in a PDA, the bin is located on the waste collecting vehicle. The PDA-based RFID when robotic/lifting arms in the waste collector loaded onto the vehicle (truck), then the weighting system (e.g., SSC) measures the weight of each bin. The data (bin ID) is then used to calculate actual waste disposal charges for each individual household and sends it to the PDA for temporary storage after emptying each bin. In additional the security of components and customers information were considered, conversely the designed system lack prior planning waste collection scheduling route since bin is only detected when the vehicle is few meters away.

On the other hand,[13]developed a wireless sensor network with three tier architecture the system was tested into three different bin samples, several sensors were involved temperature, humidity and weight sensors, ZigBee, author contented that data from the sensor nodes are transmitted to the servers through GSM/GPRS. The control station with the database server continuously receive analyze the data and update the specific bin information, the developed web based programs run in the server to facilitate the management of data on the database and bin status monitoring purpose, also the user can monitor the bin status using a web browser. Minimum energy consumption less operation cost by avoiding GPRS in every bin were also thought out, but ignoring the part for citizen participation.

Moreover, Optimal routing for efficient municipal solid waste transportation involving Geographical information system "ArcGIS" and Dijkstra algorithm for solving shortest path problem was done by [14].Authors were able to reduce the waste collection travelled distance by 9.93% of the 13 identified and selected wards,despite of the optimal route achievement also the time and cost was reduced. However, the authors didn't expound on how all type of waste stakeholders will be involved and the real time routing basis but authorities could use the solution to reduce cost and improve management.

In addition,[8]proposed a model for multi agent simulation and GIS to abstract intelligent decision support for waste management. GIS maps representing domains are abstracted to 2D lattices in order to couple with the multi agent system. Waste bins are represented as points, and actual distances are converted to appropriate units, agent follows a route on the 2D lattice, collects waste bins up to the specified time and capacity limit. Yet the simulated design didn't not focus on wireless sensor network for real time waste status detection before the collection status, moreover author didn't reflect how the household found on the map are involved with the waste management.

In general, this paper suggests the improvement for solid waste management by allowing citizens as major stakeholder of waste management to be involved during the operation. Real time monitoring and continuous reporting of waste status for wellbeing of the environment together with the prior planning for waste collection scheduling.

Table 1: Summary of related work with their advantages and limitations

| s/n | Related Literature | Citizen Involvement | Real Time Monitoring | Scheduling and optimization |
|---|---|---|---|---|
| 1. | [1] | NOT | YES | YES |
| 2. | [10] | NOT | YES | NOT |
| 3. | [11] | NOT | YES | NOT |
| 4. | [12] | YES | NOT | NOT |
| 5. | [15] | NOT | YES | YES |
| 6. | [8] | NOT | NOT | NOT |
| 7. | [14] | NOT | NOT | YES |

# 3. METHODOLOGY

## 3.1 Requirement Analysis

This work used the following software to model and simulate the real time waste collection, monitoring and generation of optimal paths for waste collections vehicles: Netlogo is an open source software, and a programmable modeling environment for simulating natural and social phenomena [16]. Netlogo 5.3.1 Multiagent Platform has been used to model and simulate the real time waste collection and generation of the optimal paths for collection. While UML diagrams were designed by using Microsoft Visio 2016 see decision logic Fig.4 and Fig.5, the general proposed architecture Fig.1 and central system architecture Fig. 2.

## 3.2 Functional Requirements

The Functional requirements of the simulated multiagent based model are highlighted as follows;

1. Waste level in a bin shall be generated per simulation tick in minutes
2. Individual bin waste levels shall continuously be updated and recorded
3. Citizen shall pay the cost per waste unit to be collected
4. The truck shall follow the generated optimal path for waste collection
5. Bin and truck shall have the maximum capacity for carrying waste load while keeping the nonfunctional performance of the simulated model to have appropriate response time, and effective throughput during the waste collection.

## 3.3 System Design

The design of the model for smart waste monitoring system has been split up into two sections which are conceptual designs and decision logic design. The sections of the designs are described in detail as follows.

### 3.3.1. Conceptual Design

*i) General architecture:* The conceptual design gives an overview of the proposed general architectural implementation and the overview of the proposed central system architecture. As shown in Fig. 1, the general architecture reveals three actors; the citizen, system administrator and truck driver. The citizen buys smart bin from the company and registers it into the central system. The bin is located at the citizen intended place (home, church, mosque, hospital, school etc.). The citizen can access information and pay for waste collection through the web. The cover of the bin will be embedded with the integrated circuit board for continuous and real time waste monitoring purposes. The Arduino Wi-Fi Shield with GSM/GPRS connection is used as the gateway for transmitting waste status data into the central system database. The administrator manages citizens registration and payment information; also browses and retrieves the processed information for waste collection and transportation then assigns the optimal route to the trucks drivers.

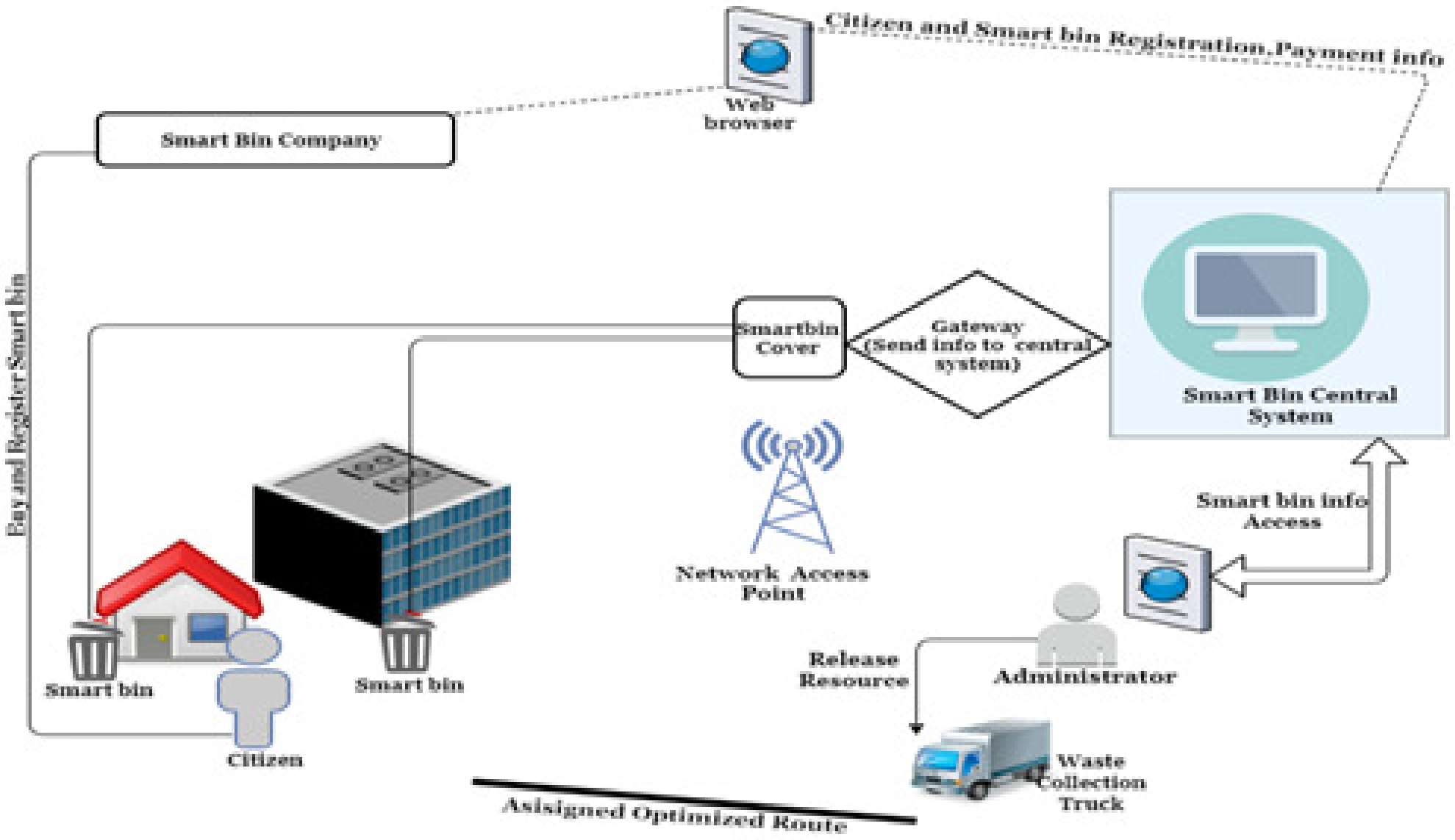

Figure 1. General architectural implementation

*ii) Central system architecture*: Fig.2 shows the central system, which is composed of three tier architectures: upper, middle and lower tiers. The upper tier comprises of a central database allied with the optimization model; the middle tier contains the gateway; and the lower tier consists of the sensor nodes. The central system receives updates, and stores the waste status via the gateways from different citizen's location after establishing connection with the server. Data are analyzed and optimal path for waste collection is established via the linked optimization model. The user graphical interface is provided with role based restrictions for both administrator and citizen from the central system.

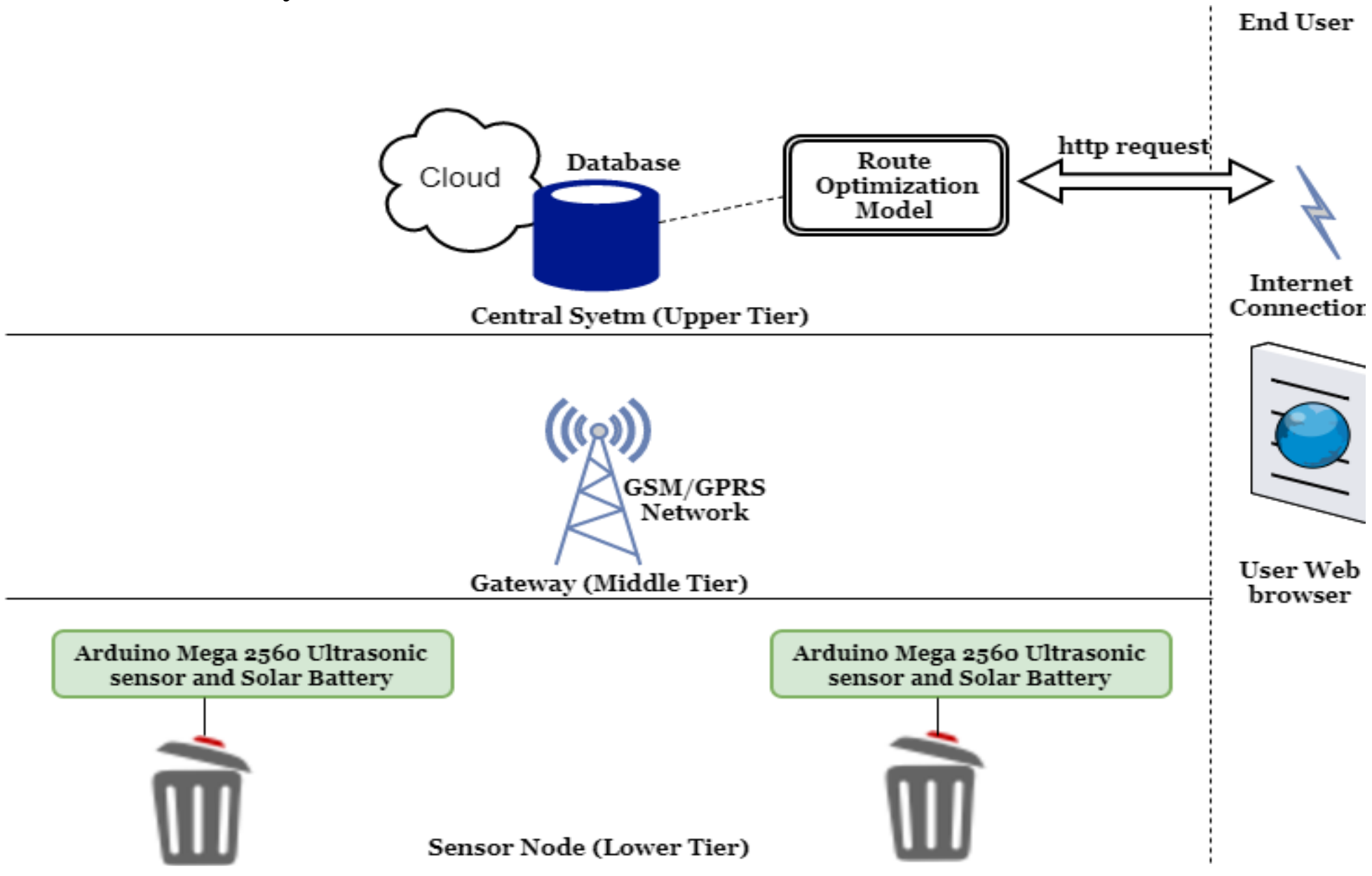

Figure 2.Smart bin and Central systems architecture.

### 3.3.2. Decision Logic

The activity diagrams Fig.3 and Fig.4 describes the building blocks and logical decision involved from bin status recording to waste bin emptying. The diagrams provide an impression on how the multi-agent model will achieve real time monitoring and collection of solid waste. Activity diagrams consist of activities, states and transitions between activities and states [17]. These are essentially flowcharts that can be used to model dynamic aspect of a system. Simulation decision logic flow charts describe the working principles of the proposed architectures in Fig.1 and Fig.2 for IoT and WSN smart waste management embracing the principles of Multi-agent based models.

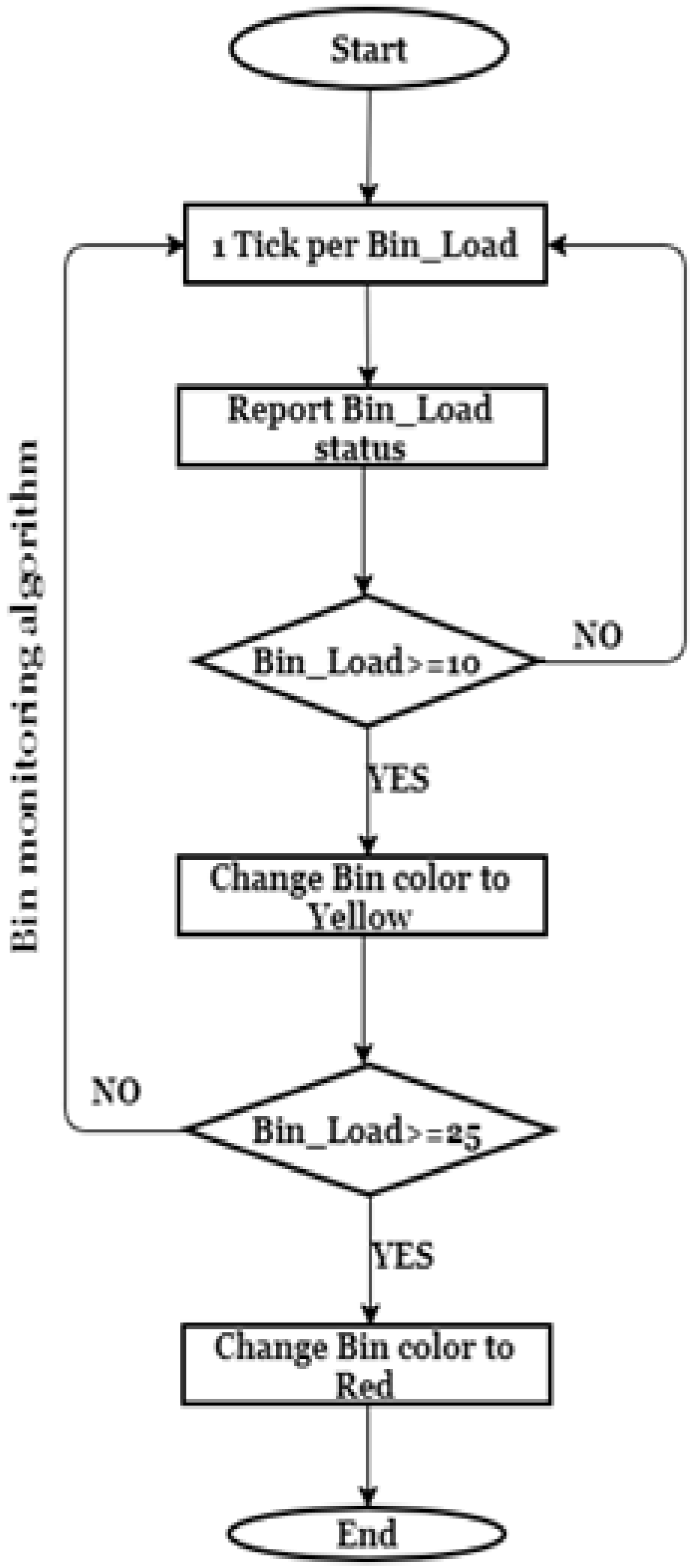

Figure 3.Bin monitoring decision logic

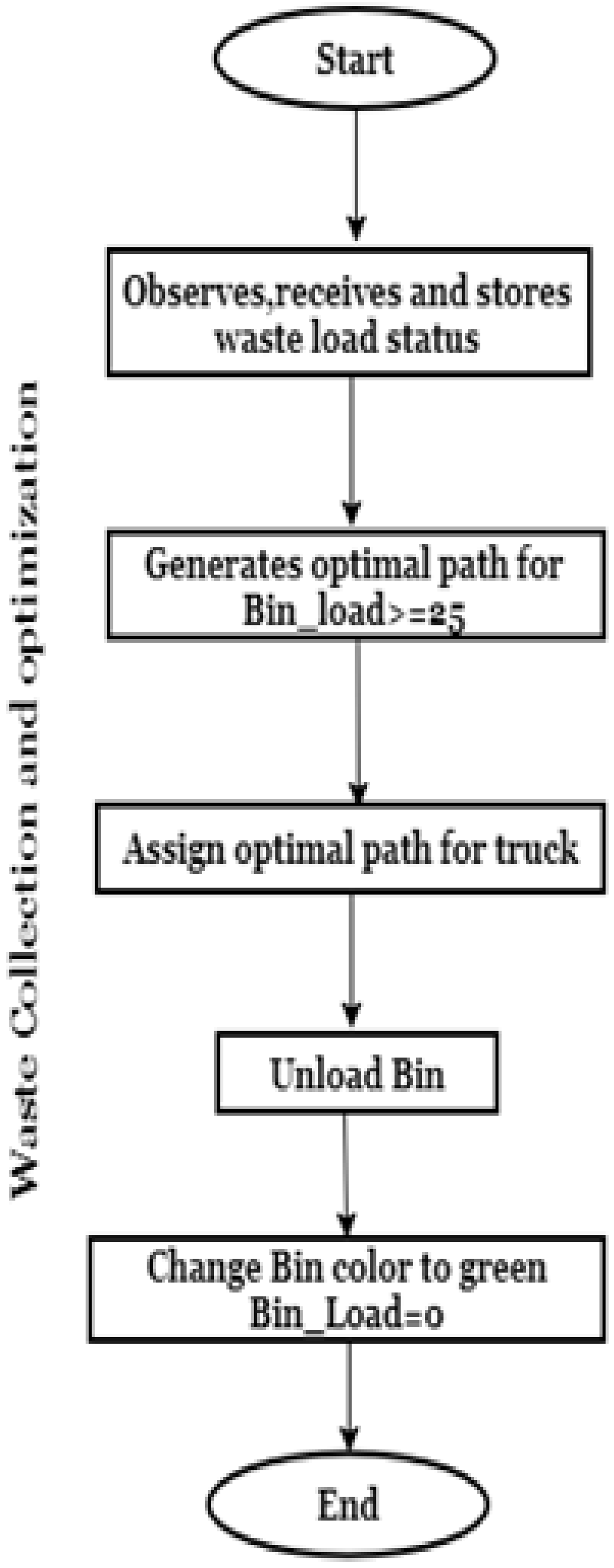

Figure 4. waste collection decision logic

# 4. SIMULATION SETUP

The Multiagent model for real time waste monitoring and collection define three main agents(turtles breed) bin, truck and citizen with their attributes and procedures, the patches form the world of a city where, smart bins were randomly placed,smart bins data extracted into .cvs format from the Netlogo system.

**Agent Bin**

The Agent bin has capacity parameter with the maximum level of 25 waste units. In this simulation 25 bins are distributed in a city following the decisions rule defined in Fig.3. Initially, all bins are green and the level of the [bin_load = 0]. The levels are increased per 1 tick when simulation time starts [bin_load = bin_load + 1], when the [bin_load>=10] the color of the agent bin changes to yellow as the first alert, again when the level [bin_load>=25] it is assumed to be the maximum capacity of the bin no more level to be added so the agent [stop] color changes to red, the levels of the bin load are reported continuously and can be exported for analysis. The simulation model's monitor provides total number of distributed bins and when they are full the number of [full_Bins uncollected] is reported, the [full_Binsuncollected] shoulddecrease each time a truck is assigned into that path for collection, else the delay in collation is observed Fig.10.

**Agent Truck**

The agent truck follows the optimal path for waste collection to the full bins ([bin_load>=25] and color set red]). The truck is simulated with the maximum capacity of [truck_load>=100] waste units, the unloaded bins by a truck returns to green and level becomes zeroas shown in Fig.4. When [truck_load>=100] is emptied into a dump and the truck load return to zero. To reduce the delay in collections number of trucks can be added. Cost per trip can be calculated by identifying the distance between points of collections together with the defined fixed cost for the operations i.e. fuel, wages and the load carried by the truck. Although for the simulation purposes this cost is constant ***(1).***

**Agent Citizen**

A citizen is another agent defined in the model in which, movement of the citizen portrays their activities which produce waste in a city. The citizen is involved in the simulated system by paying the price per unit waste they produce and collected from their own installed bins. Thus, total revenue for waste collection operation is given by:

$$R_i = N_i * p \quad \text{(i)}$$

Where $R$ is the Revenue, $N_i$ is the weight of the full bin collected, and $p$ is the price per unit weight of waste in units of currency (UC).

# 5. WASTE COLLECTION ROUTE OPTIMIZATION

In order to reduce the operational cost for waste collections, route optimization is also considered in this work. The route optimization is achieved by using Dijkstra's algorithm[18]of the Netlogo Thus, for the set of vertices Q [B, C, D, E……T] inFig.5 the algorithm identifies the vertices (place or nodes) to be visited. The edges link the start and end of vertices and the shortest path can be drawn.

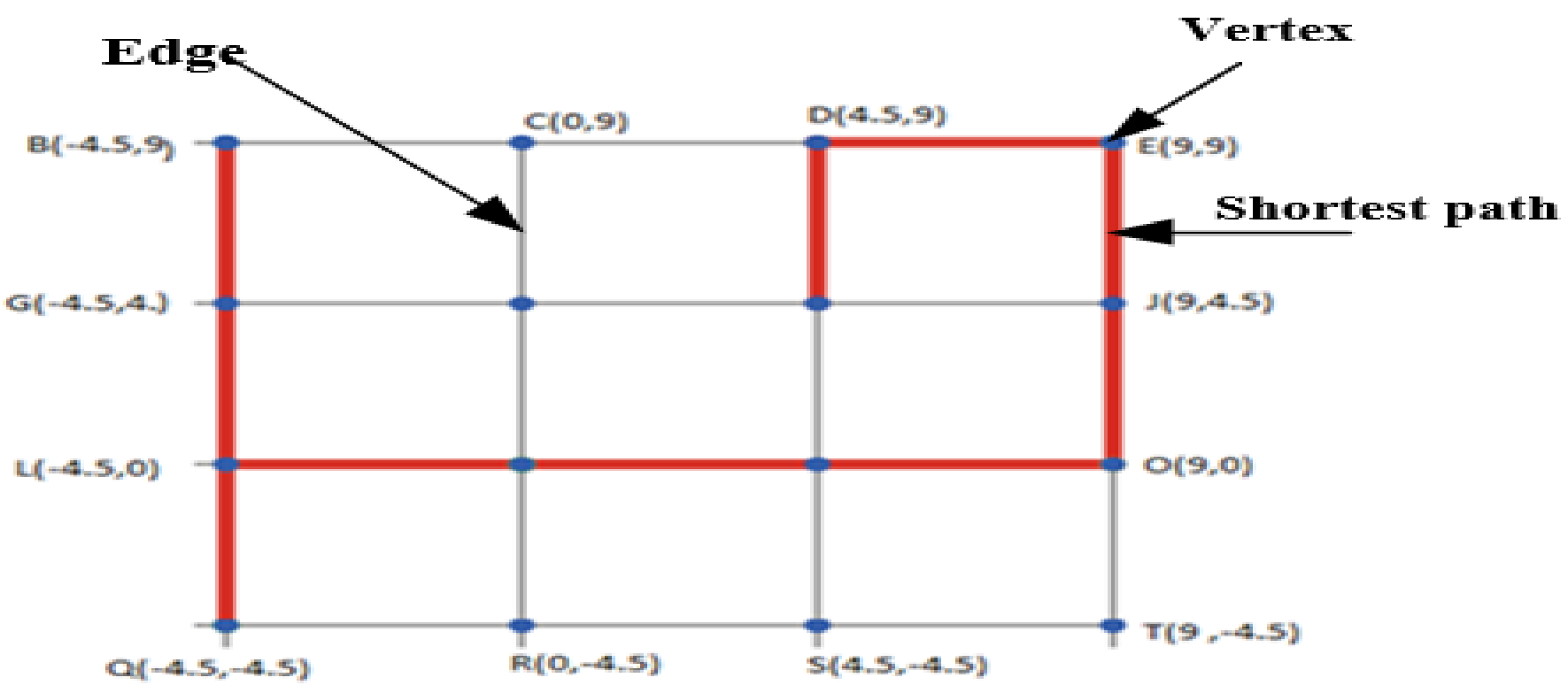

Figure 5. Set of vertices and edges linked by Dijkstra's algorithm.

## 6. RESULTS AND DISCUSSION

The implementation of Multi-agent model for real time waste monitoring and optimal route for waste collection using the Netlogo platform was verified by running several simulation cases at different simulations time T in minutes. Initially the model presents 25 bins located randomly in a city, with the initial state of [bin_load=0] and color = green. The monitors provide the number of located bins and show no number of full_bins. Parameters of bins, trucks and cost per waste unit to be paid by the citizen are also shown in Fig.6.

Simulation time T1 = 0 minutes

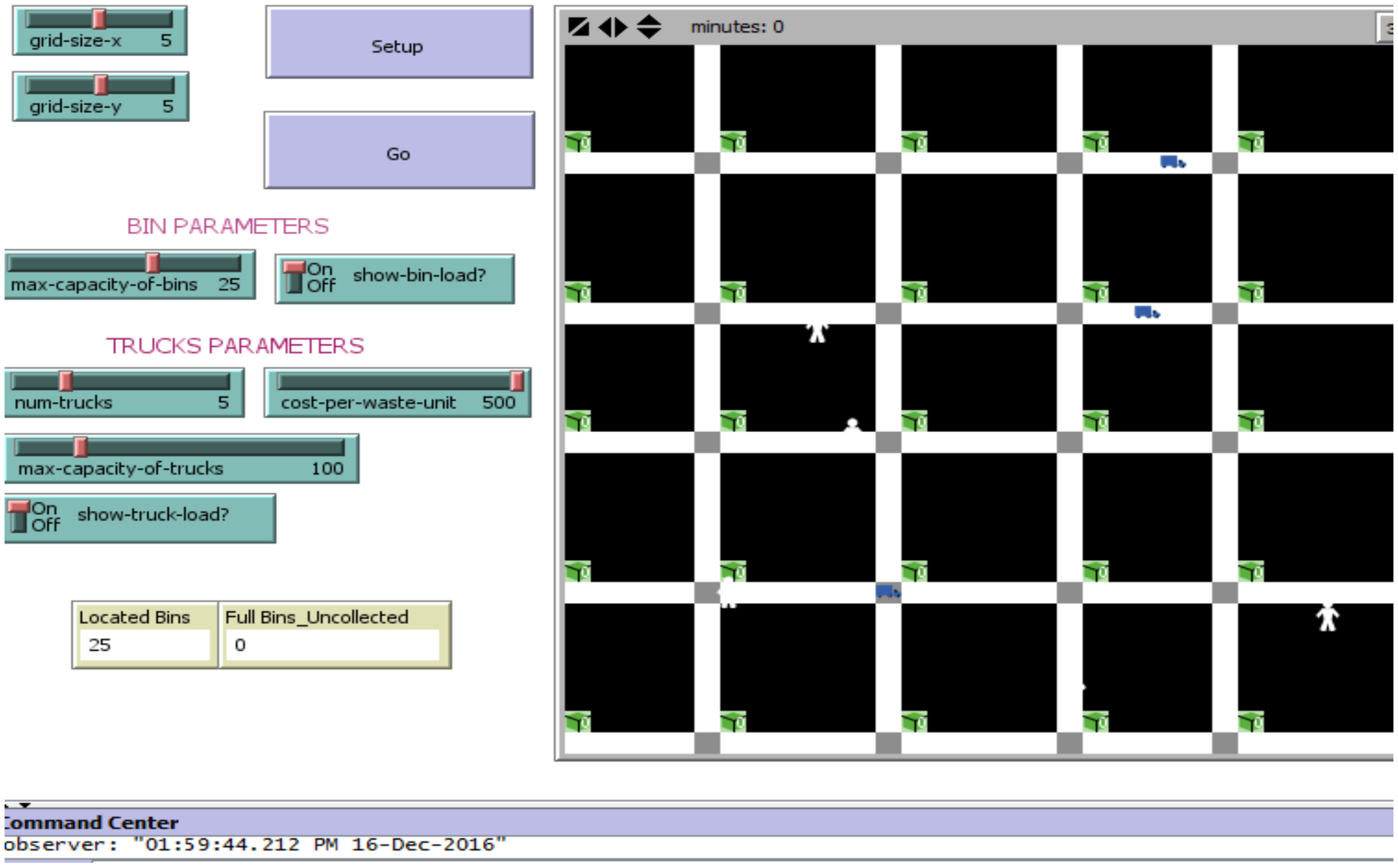

Figure 6. Initial setup for real time waste monitoring multiagent model.

The Go button run the simulation and bin levels increased per 1 tick, the bins with [bin_load>=10] set color to yellow meaning an alert that the bin is nearly to be full. The full bins are red in color [bin_load>=25].

At simulation time T2 = 53 minutes, 8 out of 25 distributed bins were full, bin number 1, 5, 6, 10,18,19,20 and 24 see Fig.7 which, also reports the continuous recording of different waste level foreach number of bin per tick, the optimal path for waste collection by a truck was obtained Fig. 8to the bins which were full. Fig.8 also shows the change in color for the different waste level inside a bin, and truck optimal path for waste collections. From this simulation case, total revenue of 100,000UCs expected to be obtained by trucks for the collection operations to all 8 full bins if 500UCsbe an assumed amount for cost per unit waste, paid by the citizen for waste collection services:

$$R_T = \sum_i^N Ni * p \quad \text{(ii)}$$

Where $R_T$ is the total revenue, $N_i$is the weight of the $i$th bin, $p$ is the price per unit weight of waste.

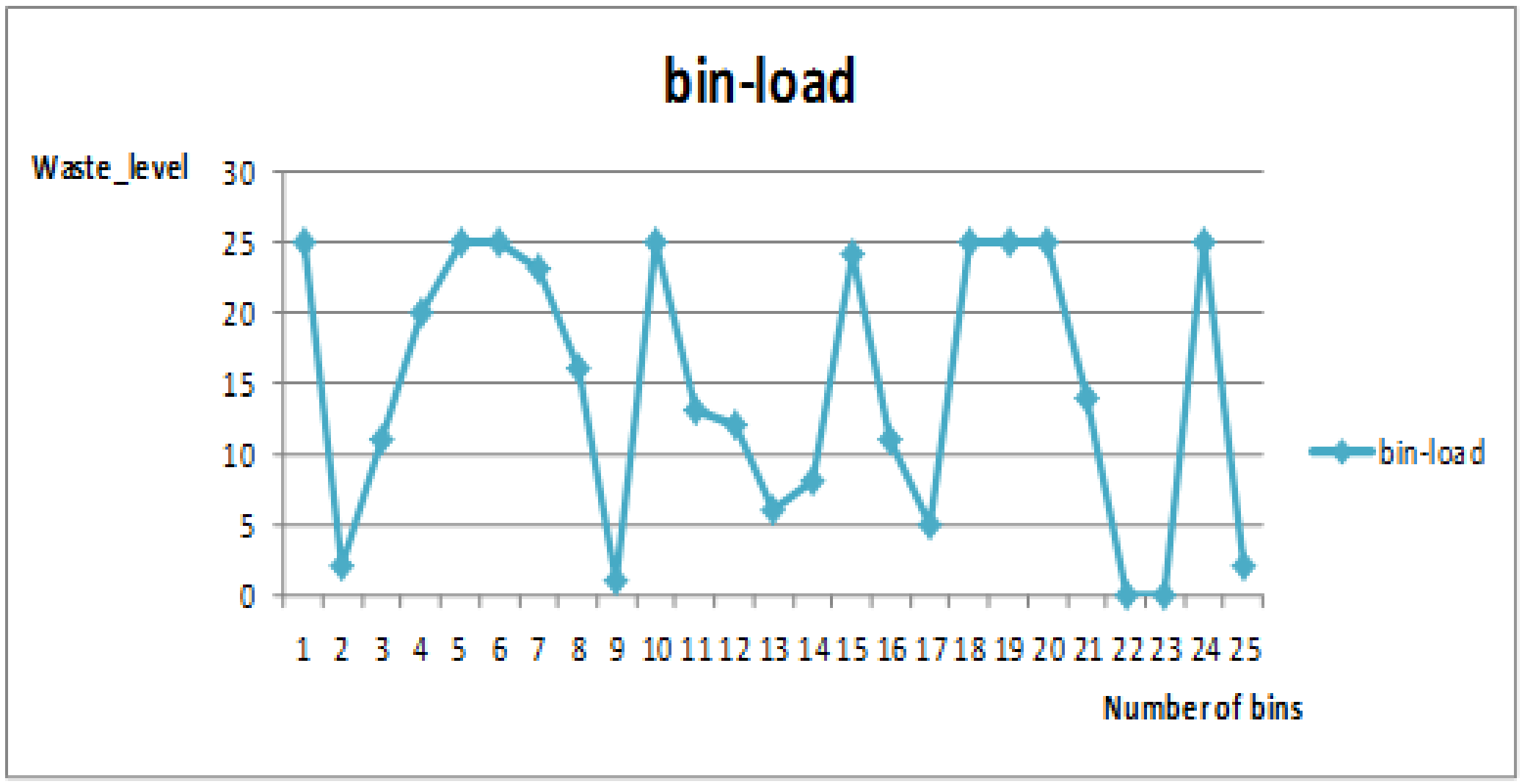

Figure 7. Bins with different waste levels at time T = 53 minutes.

At simulation time T = 93 minutes the continuous recording of different waste level for each number ofbin per tick is shown in Fig. 9. Number of uncollected waste increased to 12, new 8 full bins and other 4 which had delay in collection since T =53 minutes. The identification of vertices in the grid position for 4 bins which shown the delay in collection was done Fig.10. The obtained new optimal path included the four vertices which were uncollected from the first simulation time, vertices set for full bins with collection delay was given as;

[(D [9,9], L[(-4.5, 0)], M[(0, 0)],O[(9,0)]

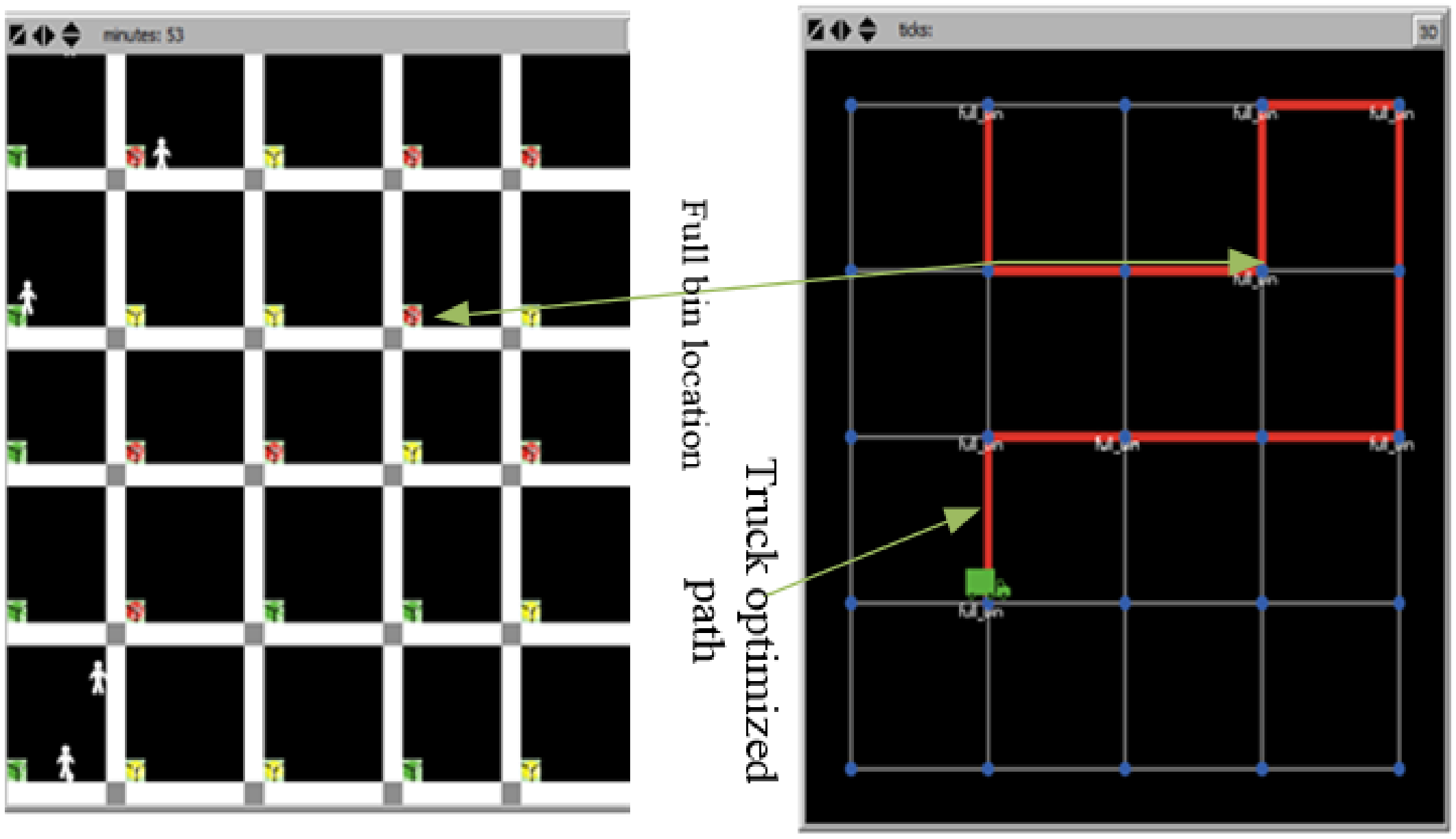

Figure 8. Color change for different bin's waste level and Truck optimal path for waste collection to 8 full bins

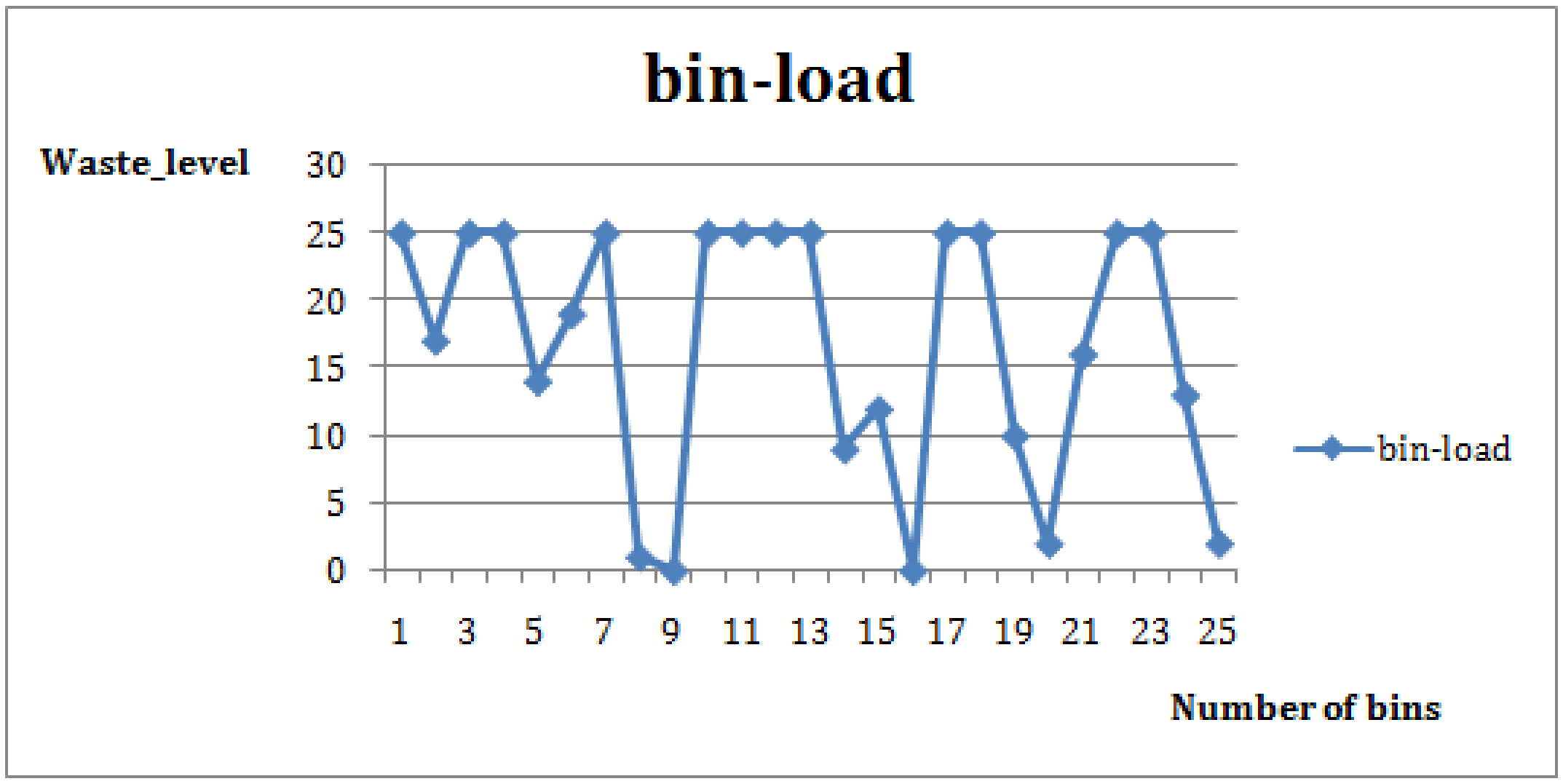

Figure 9. Bins with different waste levels at time T = 93 minutes

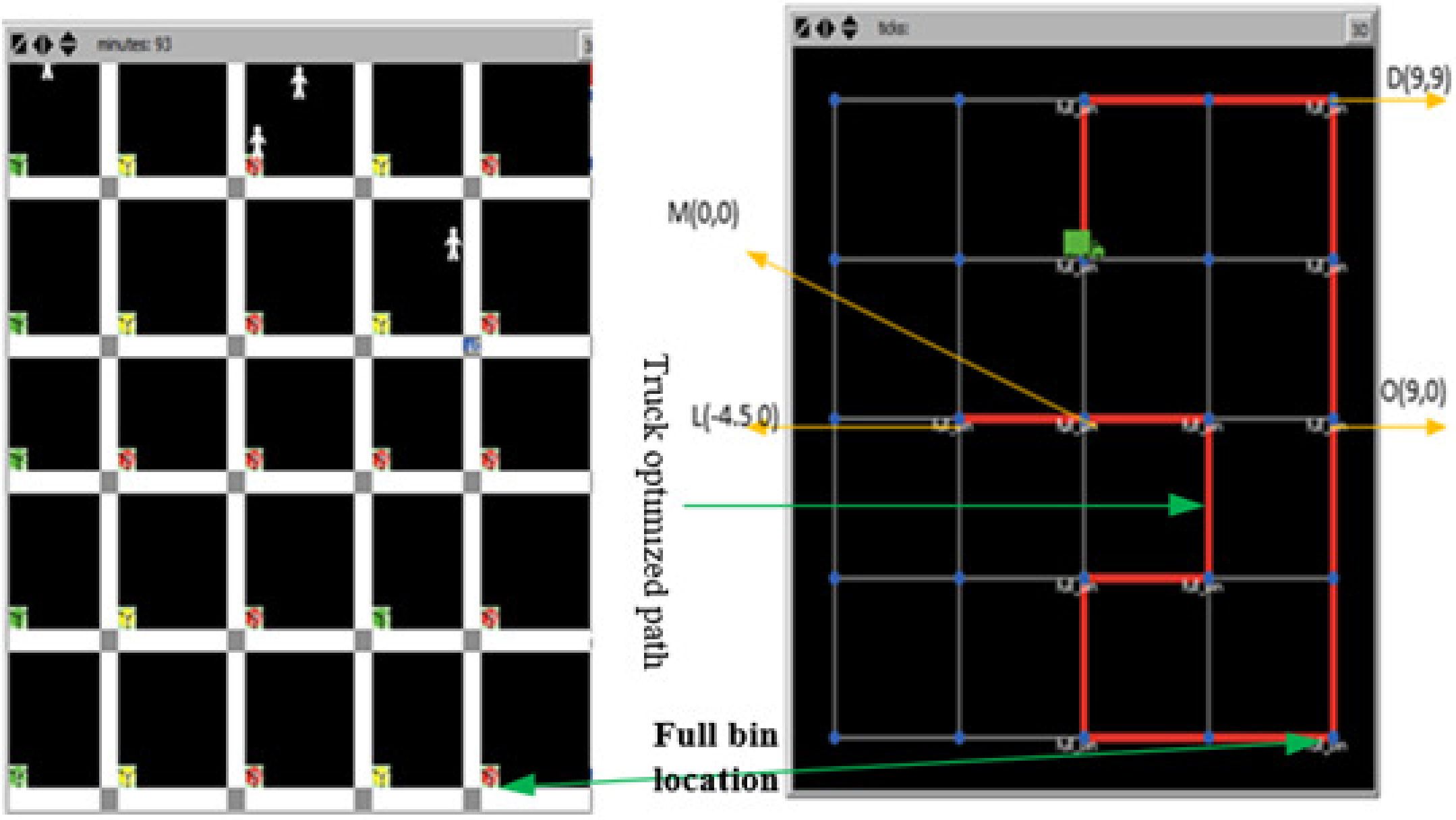

Figure 10. Identified vertices which had delay in collection and truck optimal path

## 7. CONCLUSION AND FUTURE WORK

In this paper, we have presented an overview of the proposed architectures and simulation of our on-going work towards smart solid waste management using multiagent based modeling approach. Real time and continuous reporting of waste level status from the bins ensures the clean and green environment and wellbeing of the citizen in a city. The use of WSN Technology assures efficiency of the MSW operations, despite of having technical challenges such as internet connectivity, security of components, power supply and error readings from the sensors, are to be overcome before the actual implementation.

Further work would consider full development of the WSN for real time waste monitoring and collection. Real Geographical Information System (GIS) map of the case study area will be used to identify nodes for the located smart bins. The distance from one node to another will be used to calculate cost per trip by a truck. The profit or loss of the operation will be achieved by the deduction from the collected revenues. In additional optimal path for waste collection will be achieved using the developed mathematical model for route optimization which is appropriate for supporting decision making on municipal solid waste management.

## ACKNOWLEDGMENT

The author would like to thank Ms. DevothaNyambo and Dr. Joseph Mwangoka of Department of Information Communication Science and Engineering at The Nelson Mandela African Institution of Science and Technology for the great supervision and technical guidance, and the German Academic Exchange Service (DAAD) for study fund support provided.

## AUTHORS

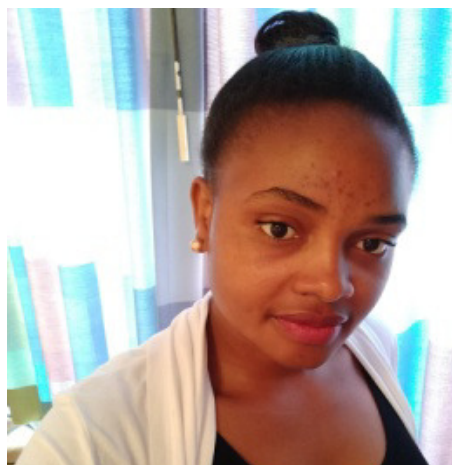

Eunice D. Likotiko is currently a master's student in Information and Communication Science and Engineering at The Nelson Mandela African Institution of Science and Technology (NM-AIST)specializing in Information Technology and System Development, also employed as Tutorial assistant at Ardhi University in the Department of Computer Systems and Mathematics. She received BSc. (Hons) degree in Information Systems Management at Ardhi University in November 2012. Her interest includes, Modelling and Simulations, Internet of Things, smart and embeded systems development.

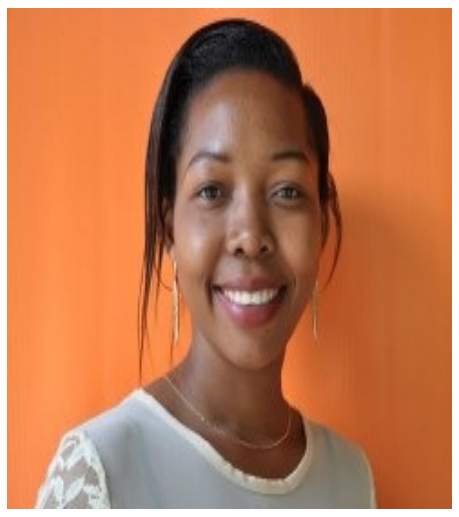

Devotha Nyambo is a PhD candidate at the Nelson Mandela African Institution of Science and Technology (NM-AIST), specializing in Information and Communication Science and Engineering. She is a prestigious fellow of the African Women in Agricultural Research and Development (AWARD) through which she has been trained in scientific skills, leadership and mentorship. Devotha's research focus is on livestock informatics. Currently, she is using Multi-Agents Research and Simulation to understand evolvement of smallholder dairy farmers. She has developed strong capabilities in paperless data collection (tools development, database designs, data management, and field work), Data analysis (descriptive statistics and modeling) and e-documents design and development for business collaboration

Dr. Joseph W. Mwangoka, is Lecturer at School of Computational and Communication Science and Engineering, at The Nelson Mandela African Institution of Science and Technology, Arusha, Tanzania. E-mail: joseph.mwangoka@nm-aist.ac.tz